\documentclass[sigconf]{acmart}
\usepackage[utf8]{inputenc}
\usepackage[T1]{fontenc}
\usepackage{array,amssymb,amsmath,graphicx,comment,multirow,microtype,makecell}
\usepackage{booktabs,color,xspace,balance,csquotes,tabularx,supertabular}

\newcommand{\ra}[1]{\renewcommand{\arraystretch}{#1}}
\renewcommand\footnotetextcopyrightpermission[1]{}

\definecolor{applegreen}{rgb}{0.55, 0.71, 0.0}

\newcommand{\para}[1]{{\vspace{7pt} \bf \noindent #1 \hspace{7pt}}}

\newenvironment{packed_enumerate}{
\begin{enumerate}
  \setlength{\itemsep}{2pt}
  \setlength{\parskip}{0pt}
  \setlength{\parsep}{0pt}
  \setlength{\topsep}{2pt}
  \setlength{\itemindent}{0pt}
}{\end{enumerate}}

\newcommand{\eg}{e.g.,\ }
\newcommand{\etal}{et~al.\xspace}
\newcommand{\ie}{i.e.,\ }

\newcommand{\fulltitle}{Ad Delivery Algorithms:\\ The Hidden Arbiters of Political Messaging}
\newcommand{\fullkeywords}{information retrieval; fairness;}

\definecolor{linkColor}{RGB}{6,125,233}
\definecolor{left}{RGB}{100,149,237}
\definecolor{right}{RGB}{205,92,92}
\hypersetup{%
  pdftitle={\fulltitle},
  pdfauthor={},
  pdfkeywords={\fullkeywords},
  bookmarksnumbered,
  pdfstartview={FitH},
  colorlinks,
  citecolor=black,
  filecolor=black,
  linkcolor=black,
  urlcolor=linkColor,
  breaklinks=true,
  hypertexnames=false
}

\begin{document}

\fancyhead{}

\renewcommand{\sectionautorefname}{\S}
\renewcommand{\subsectionautorefname}{\S}
\renewcommand{\subsubsectionautorefname}{\S}

\title{\fulltitle}

\author{Muhammad Ali$^*$}
\affiliation{\institution{Northeastern University}}
\email{mali@ccs.neu.edu}

\author{Piotr Sapiezynski$^*$}
\affiliation{\institution{Northeastern University}}
\email{p.sapiezynski@northeastern.edu}

\author{Aleksandra Korolova}
\affiliation{\institution{University of Southern California}}
\email{korolova@usc.edu}

\author{Alan Mislove}
\affiliation{\institution{Northeastern University}}
\email{amislove@ccs.neu.edu}

\author{Aaron Rieke}
\affiliation{\institution{Upturn}}
\email{aaron@upturn.org}

\begin{abstract}
Political campaigns are increasingly turning to digital advertising to reach voters. 
It is predicted that, during the 2020 U.S. presidential elections, 28\% of political marketing spending will go to online advertising, compared to 20\% in 2018 and 0.2\% in 2010.
Digital advertising platforms' popularity is partially explained by how they empower advertisers to target messages to platform users with great precision, including through inferences about those users' political affiliations.
However, prior work has shown that platforms' ad delivery algorithms can selectively deliver ads within these target audiences in ways that can lead to demographic skews along race and gender lines, often without an advertiser's knowledge.

In this study, we investigate the impact of Facebook's ad delivery algorithms on political ads. 
We run a series of political ads on Facebook---one of the world's largest advertising platforms---and measure how Facebook delivers those ads to different groups, depending on an ad's content (\textit{e.g.}, the political viewpoint featured) and targeting criteria.
We find that Facebook's ad delivery algorithms effectively differentiate the price of reaching a user based on their inferred political alignment with the advertised content, inhibiting political campaigns' ability to reach voters with diverse political views.
This effect is most acute when advertisers use small budgets, as Facebook's delivery algorithm tends to preferentially deliver to the users who are, according to Facebook's estimation, {\em most} relevant.
Moreover, due to how Facebook currently reports ad performance, this effect may be invisible to political campaigns.

Our findings point to advertising platforms' potential role in political polarization and creating informational \textit{filter bubbles}. 
We show that Facebook preferentially exposes users to political advertising that it believes is relevant for them, even when other advertisers with opposing viewpoints may be actively trying to reach them.
Furthermore, some large ad platforms have recently changed their policies to restrict the targeting tools they offer to political campaigns; our findings show that such reforms will be insufficient if the goal is to ensure that political ads are shown to users of diverse political views. 
Counterintuitively, advertisers who target broad audiences may end up ceding platforms even {\em more} influence over which users ultimately see which ads, adding urgency to calls for more meaningful public transparency into the political advertising ecosystem.

\noindent\rule{2cm}{0.4pt}

\noindent {\footnotesize $^*$ These two authors contributed equally.}
\end{abstract}

\maketitle
\pagenumbering{arabic}
\section{Introduction}
Political campaigns spend millions of dollars on ads to get their message out to voters.
Recently, much of that spending has migrated from traditional broadcast media (e.g, television, radio, and newspapers) to the Internet in the form of digital advertising.
A recent study~\cite{PoliticalAdSpending} predicts that during 2020, political ad spending overall will top \$9.9B, with over \$2.8B of that being paid to digital ad platforms.

Facebook, one of the world's largest advertising platforms, earns a significant portion of this revenue from candidates at the national, state, and local levels. The company has a dedicated site for political campaigns~\cite{FacebookPolitics} which states~\cite{FacebookPoliticsGettingStarted}:
\begin{displayquote}
Facebook advertising can help you deliver a message directly to constituents so you can better understand and engage with them on issues they care most about.
\end{displayquote}
According to Facebook's Ad Library~\cite{FacebookAdLibrary}, political campaigns have spent over \$907M on Facebook ads worldwide since May 2018, with the Trump campaign alone currently spending over \$1M each week~\cite{Trump1MWeek}.
Furthermore, a recent study found that, at the state level, "more than 10 times as many candidates advertise on Facebook than advertise on TV"~\cite{fowler-2019-advertising}.  
This adoption reflects the fact that social media platforms have substantially lowered the cost of advertising, expanding the number of campaigns who can feasibly reach voters through digital channels~\cite{fowler-2019-advertising}. 
Given the growing importance of online ads to the political discourse, it is critical to understand how complex ad platforms like Facebook operate in practice.

Much attention has been had to the {\em ad creation and targeting} phase, where the advertiser selects their desired audience and uploads their ad creative. 
Researchers have shown that advanced targeting features on ad platforms can be used to prevent certain ethnic groups from seeing ads~\cite{FacebookExcludeRace,speicher-2018-targeted}. 
For example, in 2016, the Trump campaign used these techniques to carry out "major voter suppression operations" aimed at lowering turnout among young women and black voters~\cite{TrumpMicrotargeting}, and there is evidence that Russian organizations used these tools interfere with 2016 U.S. presidential elections~\cite{MuellerMicrotargeting,ribeiro-2019-microtargeting}.

However, there is a less well-understood aspect of modern ad platforms that may be playing an equally important role:  The algorithms that ad platforms use to decide which ads get delivered to which users.
Recent work~\cite{ali-2019-discrimination} has shown that an ad platform's choices during the {\em ad delivery} phase---rooted in the desire to show {\em relevant} ads to users, ostensibly to provide users with a ``better experience''~\cite{SandbergRelevant,FacebookRelevantAds}---can lead to dramatic skew in delivery along gender and racial lines, even when the advertiser aims to reach gender- and race-balanced audiences.
In other words, an ad platform may deliver ads to a skewed subset of the advertiser's targeted audience based on the content of the ad alone, and do so unbeknownst to---and out of the control of---both the users and the advertisers.

As far as we are aware, we are the first to study whether such skews are introduced for political ads on real-world advertising platforms due to the ad delivery phase alone. 
We focus on Facebook because of its critical importance to today's digital political advertising.
We hypothesize that that Facebook may choose to deliver ads only to the subset of the political campaign's target audience that it predicts will be aligned with a campaign's views, {\em despite} attempts by the campaign to reach a diverse range of voters, and that this practice might play a role in political polarization by creating informational filter bubbles. 
Specifically, we seek to answer: {\em Is a political campaign advertising on Facebook able to reach all of the electorate?} Or, {\em is Facebook preferentially delivering ads to users who it believes are more likely to be aligned with the campaign's political views?} Additionally, {\em does Facebook vary ad pricing based on its hypothesized match between the target audience and campaign's political views?}

These questions are particularly urgent in light of the debate unfolding over the ``microtargeting'' of political ads.
In late October 2019, Twitter decided to change its policy and ban all political advertising on its platform~\cite{TwitterPoliticalAdsBan}.
In response, U.S.~Federal Election Commission (FEC) Chair Ellen Weintraub publicly argued that instead of banning such ads, ad platforms should limit political advertisers' ability to narrowly target ads to ensure that "a broad public can hear the speech and respond"~\cite{weintraub-2019-targeting}.
Shortly after, Google announced that it will significantly limit election ad targeting in order to "promote increased visibility of election ads"~\cite{GooglePoliticalAdsPolicyUpdate}.
At the time of this writing, Facebook is considering reforms of its own ad platform, but details are sparse~\cite{FacebookPoliticalAdsPolicyChange}.

Our questions regarding ad delivery are important to these developments for at least two reasons. 
{\em First}, skews resulting from ad delivery can raise fundamentally similar concerns to those raised about narrow targeting: an electorate who cannot ``hear and respond'' to political speech.
{\em Second}, ad delivery algorithms might counteract the goals of restricting microtargeting by redirecting ads according to the choices of the ad platforms (in spite of broader target audiences). Policymakers must be alert to these implications.

To test our hypotheses, we became a political advertiser and ran over \$13K\footnote{Throughout the paper we refer to prices in U.S. Dollars.} of political ads under controlled conditions, and observed how Facebook's algorithms delivered them.
Unfortunately, Facebook makes it difficult to understand ad delivery along axes of political affiliation; to measure these results, we had to design careful experiments.
{\em First}, we needed to determine {\em which users} Facebook was delivering our ads to, and whether skew (along political lines) exists among these users.
We re-used techniques published in prior work~\cite{speicher-2018-targeted,ali-2019-discrimination}, using proxies based on ground-truth data from the voter records and political donation records.
Additionally, we created audiences according to their political leaning---as inferred by Facebook---and used them simultaneously as part of an ad campaign but in a way that we can explicitly see the delivery to each subgroup.
{\em Second}, we needed to determine if it is {\em possible} for a candidate to reach their entire audience.
We used long-running ads, along with the Facebook-provided limits on how frequently a given set of ads can be shown to a user, to ``force'' the platform to consider delivering our ads to all of our targeted users.
This way, we were able to ``exhaust'' the audience to determine how much of it the platform will allow a given message to be shown to.
{\em Third}, we needed to determine how we were being charged for delivering ads to different sub-populations of the target audience.
We used Facebook's advertising reporting features, combined with proxies, in order to understand how our budget is split across users with different political leanings.

\para{Contributions}
After running our ads and analyzing the results, we present the following contributions:

\smallskip
\noindent {\em First}, we show that, despite identical targeting parameters, budgets, and competition from other advertisers, the content of a political ad alone can significantly affect which users Facebook will show the ad to.
For example, we find Facebook delivers our ads with content from Democratic campaigns to over 65\% users registered as Democrats, while delivering ads from Republican campaigns to under 40\% users registered as Democrats, despite identical targeting parameters.  
Moreover, our ``control'' ads with neutral political content that are run at the same time are delivered to a much more balanced audience (47\% Democrats), showing that preferentially delivery is a result of Facebook's ad delivery algorithm.

\smallskip
\noindent {\em Second}, we find that this effect is surprisingly not present when we target users who {\em donated} to political campaigns, rather than those who are registered for a given political party. 

\smallskip
\noindent {\em Third}, we find that that the delivery skew is present to an even greater degree when we use Facebook's own political targeting features.  
For example, when we target an audience of users who Facebook believes have ``Likely engagement with US political content (Liberal)'', combined with an equal-sized audience who Facebook considers to have ``Likely engagement with US political content (Conservative)'', we find that our ads from Democratic campaigns deliver to over 60\% liberal users (compared to ads from Republican campaigns, which deliver to 25\% liberal users).

\smallskip
\noindent {\em Fourth}, we find that it can be difficult and more expensive for political campaigns to have their content delivered to those who Facebook believes are not aligned with the campaign's views. 
For example, when re-running ads for Bernie Sanders (a liberal candidate) and Donald Trump (a conservative candidate), we find that when targeting an audience of conservative users, in the first day of the ad campaign, Facebook delivers our Sanders ad to only 4,772 users, while our Trump ad is delivered to 7,588 users.\footnote{We find a similar, but flipped, effect if we target an audience of liberal users.}
We find that the underlying reason is that our Sanders ads targeting conservative users are charged significantly more by Facebook than our Trump ads (\$15.39 versus \$10.98 for 1,000 impressions), despite being run from the same ad account, at the same time.\footnote{Again, we see a similar, flipped effect when targeting liberal users.}
Moreover, the difference cannot be attributed to some unknown underlying difference in liberal and conservative users' use of Facebook, as our neutral ad targeting the same audiences and run at the same time is delivered much more uniformly, reaching 6,822 liberal and 6,584 conservative users at a cost of \$12.07 and \$12.65 for 1,000 impressions, respectively.

\smallskip
\noindent {\em Fifth}, we find that when an ad creative and landing page shown to the users is neutral, but we ``trick" Facebook's algorithm into believing the ad leads to a page with content taken from a particular candidate's campaign web site, the skews in delivery and differential pricing are also present.
This suggests that the ad delivery decisions made by Facebook are not driven exclusively by user reactions to the ad (as all such ads appeared identical to the users), but instead are made at least partially {\em a priori} by Facebook itself.

\smallskip
Taken together, our results indicate that Facebook preferentially shows users political ads whose content it predicts are aligned with their political views, with negative implications for both users and campaigns.
For users, such delivery limits users' exposure to diverse viewpoints and---if Facebook's inference is incorrect---may pigeonhole them into a particular slice of political ads.
For campaigns, such delivery may inhibit them from reaching beyond their existing ``base'' on Facebook, as getting ads delivered to users the platform believes are not aligned with their views may become prohibitively expensive.
Importantly, these effects may be occurring without users' or campaigns' knowledge or control.

Stepping back, our findings raise serious concerns about whether Facebook and similar ad platforms are, in fact, {\em amplifying} political filter bubbles by economically disincentivizing content they believe are not aligned with users' political views. 
Put simply, Facebook is making decisions about which ads to show to which users based on its own priorities (presumably, user engagement with or value for the platform). But in the context of political advertising, Facebook's choice may have significant negative externalities on political discourse in society at large.

\para{Ethics}
All of our experiments were conducted with careful consideration of ethics.
{\em First}, we obtained Institutional Review Board review of our study, with our protocol being marked as ``Exempt''.
We did not collect any users' personally identifying information from Facebook, and did not collect any information about users who visited our site after clicking on our ads. 
{\em Second}, we minimized harm to Facebook users when running our ads by only running ``real'' ads, \ie if a user clicked on one of our ads, they were brought to a real-world page not under our control that was relevant to the topic of the ad.
In the few cases where the ads pointed to a domain we controlled, the visiting users were automatically and immediately redirected to a real page that we did not control.
{\em Third}, we minimized harm to Facebook itself by participating in their advertising system as any other advertiser would and paying for all of our ads.
We registered as an advertiser in the area of ``Social Issues, Elections or Politics''~\cite{FacebookPoliticalAdvertiser}, meaning our ads were subject to the same review as the ads of other political campaigns.  
{\em Fourth,} we minimized the risk of altering the political discourse through careful choices of the ad content (Section~\ref{sec:ad-copies}), and running approximately the same number of copies of ads for Republican and Democratic candidates, with the same budgets. 
The total amount we spent on political advertising while collecting data for this paper (\$13.7K) is minuscule compared to the ad budgets of real campaigns in the same period (likely in the millions of dollars~\cite{FacebookAdLibrary}).

\para{Limitations}
It is important to note the limitations of our study (see Section~\ref{sec:limitations} for a detailed discussion).
Most importantly, we can only report results of {\em our own ads}; we are unable to make any statements about the extent to which any effects we observe exist for political ads run by real political campaigns, or political ads {\em in general}.
However, the fact that we observe statistically significant skews in our small set of ads suggests that the effects we observed are likely present in the delivery of other political ads as well.

\medskip

The rest of this paper is organized as follows.
Section~\ref{sec:background} provides background on Facebook's advertising platform and Section~\ref{sec:related} details related work.
Section~\ref{sec:methodology} gives an overview of our methodology, and Section~\ref{sec:experiments} presents the results of our experiments.
Section~\ref{sec:discussion} offers a concluding discussion.

\section{Background}\label{sec:background}
In this section, we provide some basic background about Facebook's advertising platform, the subject of this study, necessary to understand our methodology (described in Section~\ref{sec:methodology}).

\subsection{Overview}
On Facebook, like many ad platforms, there are two phases to advertising: {\em ad creation} and {\em ad delivery}.

\para{Ad creation} Ad creation refers to the process by which an advertiser submits their ad to the Facebook ad platform.
During this stage, advertisers provide the contents of the ad (the images, videos, text, and destination link collectively called the {\em ad creative}), and specify the {\em target audience} of users on the platform to whom they wish the ad to be shown (see \S\ref{subsec:targeting}).
Advertisers often run many ads that are related; collectively, these are called an {\em ad campaign}.
Before submitting their ad campaign, advertisers also specify an {\em objective} (i.e., what they want to achieve with the campaign, see \S\ref{subsec:objective}) and the {\em ad budget} they are willing to spend to achieve that objective (see \S\ref{subsec:budget}).
The ads then enter a review process to be approved to run on the platform (see \S\ref{subsec:review}).

\para{Ad delivery} Ad delivery refers to the process by which the Facebook ad platform selects which ads get shown to which users.
Before displaying an ad to a user, the platform will hold an ad auction to determine which ad, from among all ads that user is eligible to see (by virtue of their inclusion in ads' target audiences), will be shown (see \S\ref{subsec:delivery}).
While the ad campaign is active, the platform provides a semi-live, detailed breakdown to advertisers of how their ads are being delivered (see \S\ref{subsec:reporting}).

\subsection{Targeting}\label{subsec:targeting}
There are many different ways for advertisers to target ads, ranging from users' demographics and interests to their personally identifiable information (PII)~\cite{speicher-2018-targeted,venkatadri-2018-targeting}. 
We briefly describe these below.

\para{Detailed targeting}
Facebook pioneered, and is continuing to aggressively market, ways for advertisers to target its users via {\em user attributes}~\cite{speicher-2018-targeted}.
These attributes cover a variety of aspects of users' lives, ranging from demographic features to online activity and even offline information, often acquired without users' explicit consent or knowledge.
In the context of politics, Facebook derives attributes that indicate whether users are ``interested in'' various political candidates (e.g., Donald Trump, Elizabeth Warren), as well as more general attributes about user behaviors, such as ``Likely engagement with US political content (Conservative)" or ``Likely engagement with US political content (Liberal)".
The exact methodology by which Facebook infers such attributes is not disclosed, but likely involves profile data provided directly to Facebook, data from activity on Facebook (e.g., ``Liking'' Pages or explicit or implicit patterns of interaction with particular content), inferences based on attributes of a user's friends, and data inferred from users' behavior off of Facebook.\footnote{Recall that Facebook receives information from a variety of sources beyond the Facebook website and app, including Facebook Pixel tracking~\cite{FacebookPixelAudience}, app data sharing~\cite{FacebookMobileAdIds}, third-party data brokers~\cite{venkatadri-2019-databrokers}, and location data~\cite{faizullabhoy-2018-fbattacks}.}

\para{Custom audiences} 
Facebook also allows advertisers to target users directly using their PII via a tool called {\em Custom Audiences}~\cite{venkatadri-2018-targeting}.
Using Custom Audiences, advertisers can upload up to 15 different kinds of PII to Facebook, ranging from names to email addresses to phone numbers to dates of birth. 
Facebook then matches these values against their database in order to build an audience for the advertiser; the advertiser is then allowed to target their ads to just the users who match.

\subsection{Objective}\label{subsec:objective}
When creating ad campaigns, advertisers on Facebook are asked to specify their {\em objective}, or what they are trying to achieve.
Common objectives include ``Reach" (showing the ad to as many users as possible), ``Traffic" (showing the ad to the users most likely to click), and ``App Installs" (showing the ad to the users most likely to install the advertiser's app). 

Within each objective, advertisers must also specify an \textit{optimization}, indicating how Facebook should achieve their objective. 
For the ``Reach" objective, the available optimizations are ``Reach" (the default, showing ads to as many {\em users} as possible) and ``Impressions" (showing ads as many {\em times} as possible). 
For the ``Traffic" objective, the available optimizations are ``Link Clicks" (the default, showing ads to the users most likely to click), ``Landing Page Views" (showing ads to the users most likely to click {\em and} visit the destination page), ``Daily Unique Reach" (showing ads to users most likely to click, at most per day), and ``Impressions" (showing ads to the users most likely to click, but maximizing the number of impressions).

We hypothesize that political advertisers are likely to commonly use the ``Reach" campaign objective together with default the ``Reach" optimization---representing the goal of getting their message out to as many people as possible---and the ``Traffic" campaign objective together with the default ``Link Clicks" ad set optimization---representing the goal of getting as many people as possible to visit their campaign page. 
In this paper we use the ``Traffic'' optimization in exeriments targeting registered voters and donors (see Section~\ref{subsec:contentskew}) and ``Reach'' optimization in all other experiments.

\subsection{Bidding, budgets, and billing}\label{subsec:budget}
When creating an ad, the advertiser must tell Facebook their bid, which takes the form of an {\em ad budget}.
These budgets are either a daily or lifetime budget for the ad, allowing Facebook to spend the advertiser's money between and within auctions according to an algorithm that is not publicly known.\footnote{Facebook only says: ``Facebook will aim to spend your entire budget and get the most 1,000 impressions using the lowest cost bid strategy" for the ``Reach" campaign-level objective, and ``Facebook will aim to spend your entire budget and get the most link clicks using the lowest cost bid strategy" for the ``Traffic" campaign-level objective. An optional ``Bid Control" (maximum bid in each auction) and ``Cost control" (the average cost per link click) are also available for the ``Reach" and ``Traffic" campaign-level objectives, respectively.}
When using the interface, Facebook also provides advertisers with an ``Estimated Daily Reach'', which Facebook defines as the estimated number of users who would be exposed to an ad given the targeting criteria and budget; this feature helps advertisers with an understanding of how far their budget will go for the selected audience.
Although the bid and cost control options are offered, without the knowledge of the auction, they are difficult to set effectively, and therefore, it is natural to only specify a budget and rely on Facebook to do the bidding. 

How the advertiser is actually charged for their ad campaign depends on the objective. 
For ``Reach", the advertiser is charged per impression; for ``Traffic" the advertiser is also charged per impression unless the optimization is ``Link Clicks" (in which case the advertiser can choose to be charged {\em per click} if they wish).

Finally, if the advertiser chooses the ``Reach" objective and ``Reach" optimization, they are allowed to specify a {\em frequency cap}, which allows them to set the maximum number of impressions a single user would see over a specified number of days.
We use this feature later in the paper to force Facebook to deliver our ads to many users in our audience.

\subsection{Ad review}\label{subsec:review}
Once the ad creation phase is complete, and before the ad enters the ad delivery phase, it is submitted to Facebook for review.\footnote{Most platforms have a review process (consisting of a combination of automated and manual review) to prevent abuse or violations of their advertising policies~\cite{FacebookAdReview,TwitterAdReview}.}
We observed that most of our Facebook ads were approved within 30 minutes, some spent hours in review, and a few were never approved.
The criteria and internal mechanisms for approval are not entirely clear, and precise reasons for why certain ads are rejected are not given.
In this work, we only report on experiments where all necessary ads were approved before their scheduled start time.

\subsection{Ad delivery}\label{subsec:delivery}
Ad platforms including Facebook commonly use {\em ad auctions} to select which ads to show to users.
Historically, this auction took only the advertiser's bid price into account; more recently, Facebook considers other features such as the overall performance of the ad and the platform's estimate of how relevant the ad is to the browsing user~\cite{FacebookAdAuctions,SandbergRelevant}.
Facebook says as much in its documentation for advertisers~\cite{FacebookAdAuctions}: 
\begin{displayquote}
{[}W{]}e subsidize relevant ads in auctions, so more relevant ads often cost less and see more results. In other words, an ad that’s relevant to a person could win an auction against ads with higher bids.
\end{displayquote}
Facebook explains that it measures {\em relevance} as a composite of {\em estimated action rates} ("[a]n estimate of whether a particular person engages with or converts from a particular ad" and {\em ad quality} ("[a] measure of the quality of an ad as determined from many sources including feedback from people viewing or hiding the ad")~\cite{FacebookAdAuctions}. 

In short, Facebook plays a significant role in determining which users see which ads, based on its own judgment about which ads are likely to be "relevant" to particular users, its own judgement of how to bid on an advertiser's behalf and distribute the specified budget among auctions, and possibly other considerations connected to its business interests. 
This role---in the context of political ads---is the key phenomenon that this paper seeks to explore in its experiments.

\subsection{Reporting}\label{subsec:reporting}
During the ad delivery phase, Facebook provides semi-live~\cite{ali-2019-discrimination} detailed statistics to advertisers about how their ad is being delivered.
In particular, Facebook reports the {\em impressions} (the number of times the ad was shown), the {\em reach} (the number of unique users who saw the ad), the {\em clicks} (the number of times users clicked on the ad), and the {\em spend} (how much money was spent).
Facebook also allows advertisers to obtain breakdowns of these performance metrics along a few axes, most notably gender (broken into ``Male'', ``Female'', and ``Other''), age (broken down into brackets of 10 year increments), and location (broken down into Designated Market Area~\cite{NeilsonDMARegions}, or DMA).
In other words, Facebook will tell advertisers how much they have spent on users in different regions (or of different genders, etc.), and how many clicks/impressions/reach those users represent.
Notably for this work, Facebook does not provide breakdowns along axes such as Facebook's estimated political leaning.

\section{Related work}\label{sec:related}
We now provide a brief overview of related work on skew in ad delivery, filter bubbles, and political advertising.

\para{Skew in ad delivery}
Recently, concerns have been raised about how the platforms' desire to show ``relevant'' ads to users may raise issues of lack of fairness and lack of transparency.
Recent work~\cite{lambrecht-2018-algorithmic,ali-2019-discrimination} demonstrated that on Facebook, ads can be showed to skewed subsets of the target audience, sometimes with dramatic effects on delivery (e.g., ads targeting the same audience but with different content can be shown to over 95\% women or less than 15\% women, depending only on the content of the ad and {\em not} on the advertiser's targeting choices or competition from other advertisers). 
In some cases, skews in ad delivery may be unsurprising or desirable. In others, such as civil rights areas and political ads, they can raise serious issues that demand research and, potentially, regulation.

\para{Online political advertising}
Prior work, conducted mostly in the form of laboratory experiments, indicated high efficiency of written persuasion personalized to the psychological profile and motivation of the recipient~\cite{moon-2002-personalization,hirsh-2012-personalized}.
More recently, Matz~\etal conducted a large-scale experiment in which they showed that Facebook ads tailored to individual's psychological characteristics yielded higher click-through and conversion rates compared to non-personalized ads and mismatching ads~\cite{matz-2017-psychological}.
Matz~\etal relied on the mechanism first documented by Kosinski~\etal: that personality traits of an individual can be accurately inferred from the content that they ``Like'' on Facebook~\cite{kosinski-2013-private}.
Other researchers, however, pointed out that the unknown optimization mechanisms employed by Facebook might obfuscate the measurement of effectiveness of these personalized ads~\cite{eckles-2018-field}.
Our prior results~\cite{ali-2019-discrimination} indicate that Facebook does, indeed, further refine even precisely targeted audiences, introducing demographic and political biases in the reached audiences, beyond those intended by the advertiser.

\para{Filter bubbles}
The extent, or even the existence, of the {\em filter bubble} effect has been a point of contention both in academia and in popular media.
After the initial reports by Eli Pariser~\cite{pariser-2011-filterbubble} scholars have attempted to measure the phenomenon in services including Google Search~\cite{hannak-2013-filterbubbles,flaxman-2015-filter}, Google News~\cite{haim-2018-burst}, and Facebook~\cite{bakshy-2015-exposure}.
However, the observed differences could often be explained by user location differences (in the case of Google Search) or attributed to an individual's choice of friends to follow (in the case of Facebook), rather than stemming from algorithmic personalization of an individual's experience.
Furthermore, Bail~\etal warn against overexposing users to messaging from politicians they do not support as it appears to increase, rather than decrease their partisanship~\cite{bail-2018-exposure}.

Other previous work indicated that the effect might be more pronounced in ads than in organic content.
Datta~\etal showed that personal attributes are used in ad selection, specifically that changing one's self-reported gender influences the job ads one sees~\cite{datta-2015-automated}.
More recently, Ali~\etal showed that the demographic distribution of the audience that receives the ad changes depending on the ad content, even if the same audience was targeted~\cite{ali-2019-discrimination}.

Our work provides further evidence that the filter bubble effect is pronounced in the ads users are exposed to; we show that attempting to ``burst'' the political advertising filter bubble can prove expensive, especially for smaller advertisers.
As far as we are aware, we are the first to study the filter bubble effect due to ad delivery aspects of political messaging; prior work on filter bubbles for political content focused on possibilities of disparate treatment of organic rather than sponsored content or disparate treatment during other parts of the process, such as during the ad review stage~\cite{FBKylReview, Kreiss2019pc}.

\section{Methodology}\label{sec:methodology}

In this work we aim to answer two related, but separate questions, and design our experiments accordingly.
\textit{First,} we want to verify whether the skew in delivery reported in previous work~\cite{ali-2019-discrimination} exists along the lines of political affiliation for political ads.
To this end, we replicate the study setup from~\cite{ali-2019-discrimination} as closely as possible, including setting the campaign objective to ``Traffic''.
\textit{Second,} we ask whether a political campaign determined to reach users who may not be aligned with its views---and explicitly requesting such audience from Facebook---can achieve their goal.
To be able to better answer this question, we set the campaign objective to ``Reach''. 

For the sake of clarity, we run ads for only one Democratic presidential candidate (Bernie Sanders) and compare their performance to that of the ads for only one Republican candidate (Donald Trump).
We choose these two candidates because at the time of experiment design (early July 2019), they had spent most on Facebook advertising among the major candidates of each party~\cite{FacebookAdLibrary}.
Therefore, their election performance is least likely to be influenced by ads run on our limited budget.

\begin{table*}[t!]
\centering
\ra{1.5}
\begin{tabular}{p{6.75cm}ccccccccccc}\toprule
\multirow{2}{*}{\bf DMA(s)~\cite{NeilsonDMARegions} } & \multicolumn{2}{c}{\bf CA$_A$} & \phantom{a} &  \multicolumn{2}{c}{\bf CA$_B$ } & \phantom{a} &  \multicolumn{2}{c}{\bf CA$_C$ } & \phantom{a} &  \multicolumn{2}{c}{\bf CA$_D$ } \\
\cmidrule{2-3} \cmidrule{5-6} \cmidrule{8-9} \cmidrule{11-12}
 & {\bf Dem} & {\bf Rep} && {\bf Dem} & {\bf Rep} && {\bf Dem} & {\bf Rep} && {\bf Dem} & {\bf Rep}\\
\midrule
Greensboro, Charlotte & 70,000 & {0} && {0} & {70,000} && {70,000} & {0} && {0} & {70,000}\\
\multirow{2}{*}{\parbox{6.75cm}{
Wilmington, Raleigh-Durham,\\
Greenville-(New Bern and Spartanburg)
}} & \multirow{2}{*}{0} & \multirow{2}{*}{63,137} && \multirow{2}{*}{70,000} & \multirow{2}{*}{0} && \multirow{2}{*}{0} & \multirow{2}{*}{54,000} && \multirow{2}{*}{64,166} & \multirow{2}{*}{0}\\
& & & & & & & &\\
\bottomrule
\end{tabular}
\label{stats:audience_voters}
\caption{Number of uploaded records for Custom Audiences created using publicly available voter records.  We divide the DMAs in the state into two sets, and create two audiences, each with voters registered with one party per DMA set (CA$_A$ and CA$_B$). We repeated this process with separate voter records (creating CA$_C$ and CA$_D$), allowing us to run experiments on separate audiences.  The number of uploaded records does not match, as we uploaded records so that the Estimated Daily Reach was the same.
}\label{stats:audience_voters}
\end{table*}

Next, we provide more details on the audiences and ad campaigns in our experiments, how we measured their performance over time, and the statistical apparatus necessary to interpret the results.

\subsection{Creating audiences}\label{sec:creating-audiences}
We use two mechanisms for targeting audiences on Facebook: Custom Audiences and detailed targeting.

Recall that we are interested in studying the skew in delivery along political lines.
Since Facebook does not provide ad delivery breakdowns by political leaning, but does provide breakdowns by location (Section~\ref{subsec:reporting}), we craft our Custom Audiences in such a way that the statistics about the political leanings of the actual recipients of an ad can be inferred from the statistics about their location. 
Specifically, we follow the method introduced by prior work~\cite{speicher-2018-targeted,ali-2019-discrimination}.

\begin{figure*}[t!]
\centering
\includegraphics[width=1\linewidth]{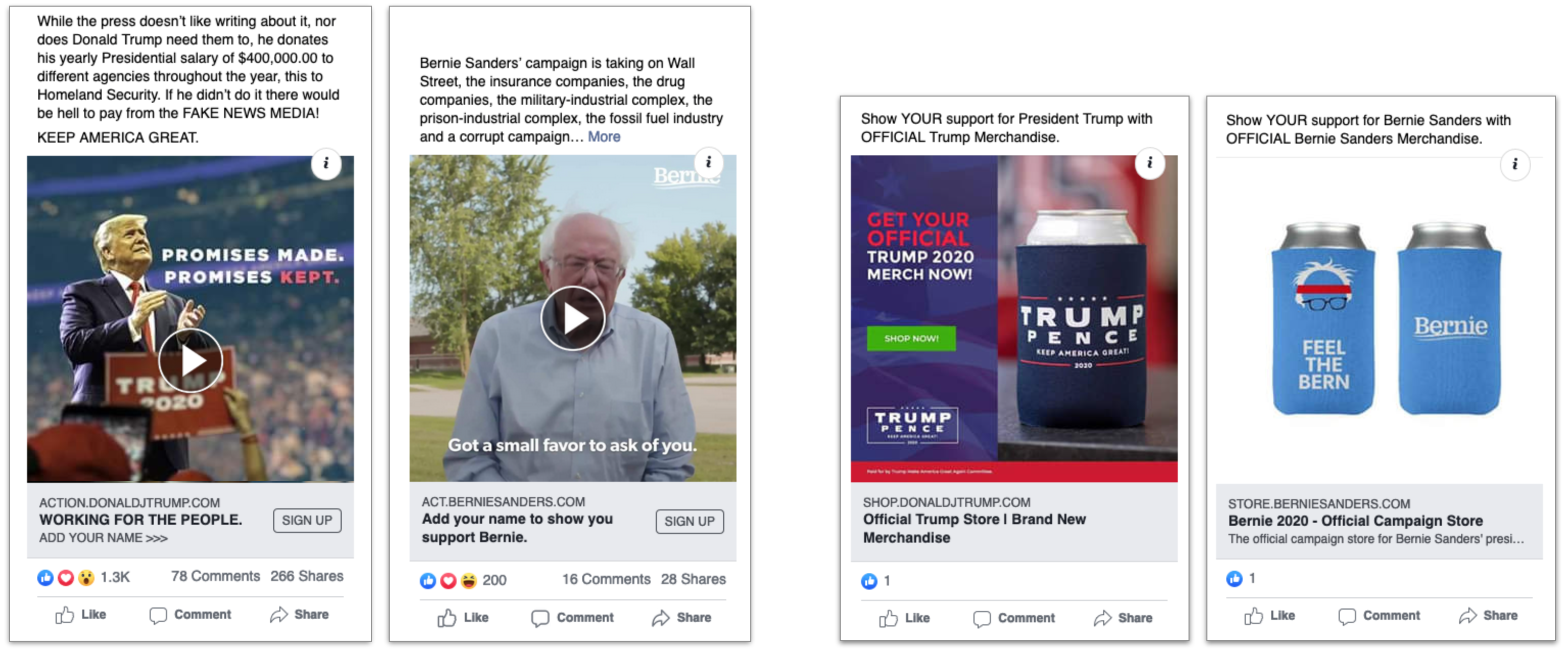}
\caption{Ads used in our experiments concerning political issues and promoting candidates' merchandise. The ads were copies of real ads run on Facebook by the official campaigns, with the exception for Bernie Sanders related merchandise (as his store has no official Facebook advertising).
}
\label{fig:ads}
\end{figure*}

\para{Custom Audiences from voter records} 
We obtained publicly available voter records from North Carolina, which, in addition to PII, include each voter's political party registration (if one exists).
We then create Custom Audiences as follows: 
\begin{enumerate}
  \item Divide the DMAs in North Carolina into two sets of roughly equal population sizes.
  \item Create a Custom Audience that contains PII of registered Democrats from the first set of DMAs and another Custom Audience with PII of registered Republicans from the second.
  \item Upload the lists to Facebook, and compare their ``Estimated Daily Reach'' statistics provided by Facebook.
  \item If one audience has a higher Estimated Daily Reach, subsample it and re-upload until the estimates match.
  \item Repeat steps (2)--(4) with the opposite assignment: registered Republicans from the first set of DMAs and registered Democrats from the second set of DMAs.
\end{enumerate}
Table~\ref{stats:audience_voters} shows the summary statistics of the audiences created from voter records.

There are two distinctions between our method and previous work~\cite{ali-2019-discrimination}.
{\em First}, we introduce an additional step in an attempt to create audiences consisting of roughly equal numbers of Democrats and Republicans.
We do so for ease and clarity of subsequent analysis; it is not strictly necessary to do so, as we will compare the delivery of two ads {\em targeting the same audience and run at the same time} to observe differences due to ad delivery (thus, any differences in population, usage, or time of day will affect both campaigns equally).
{\em Second}, Facebook no longer provides advertisers with the number of uploaded records that match Facebook users, as these estimates have been shown to leak private information about individuals~\cite{venkatadri-2018-targeting,venkatadri-2019-pii}.
Instead, we use the Estimated Daily Reach provided by Facebook with a budget set to a very high number (\eg \$1M/day), thereby obtaining an estimate of the {\em total} daily active users in the uploaded audience.

\para{Custom Audiences from donor records} 
We build separate Custom Audiences from publicly available donor records for political campaigns.
The U.S. Federal Election Commission (FEC), in particular, makes publicly available the PII of all contributors who have donated a total of \$200 or more towards a political campaign~\cite{FECContributions}.
We obtain data from the FEC for individuals who have donated to the ``Bernie 2020" or the ``Donald J. Trump for President" committees as of July 1, 2019 to craft Custom Audiences of users who actively engage with these campaigns.
Because there are fewer donors for Bernie Sanders than for Donald Trump in FEC data, we also use mid-year FEC filing by ActBlue, a popular Democratic fundraising platform~\cite{ActBlueMid2019Filing, BuzzFeedNewsFECAnalysis} to obtain a list of Democratic donors who donated less than \$200.
Since the FEC data isn't limited to a particular state or region, we randomly split all 210 U.S.~DMA regions~\cite{NeilsonDMARegions} into two sets, and then create Custom Audiences of approximately equal numbers of Democratic and Republican donors in each, by relying on the estimated daily reaches (as in Steps (2)--(4)) above.
Table~\ref{stats:audience_donors} shows the size and configuration of our audiences created from donor records.

\begin{table}[t!]
\centering
\begin{tabular}{p{2cm}ccccc}\toprule
\multirow{2}{*}{\bf DMAs~\cite{NeilsonDMARegions} } & \multicolumn{2}{c}{\bf CA$_E$} & \phantom{a} &  \multicolumn{2}{c}{\bf CA$_F$ }\\
\cmidrule{2-3} \cmidrule{5-6}\
& {\bf Trump} & {\bf Sanders} && {\bf Trump} & {\bf Sanders}\\
& {\bf donors} & {\bf donors} && {\bf donors} & {\bf donors}\\
 \midrule
 DMA Set 1 & 40,973 & 0 && 0 & 32,000\\
 DMA Set 2 & 0 & 32,000 && 41,458 & 0\\ 
\bottomrule
\end{tabular}
\caption{Overview of Custom Audiences built from public FEC donor records and ActBlue.  The number of uploaded records does not match, as we uploaded records so that the Estimated Daily Reach was the same. 
}
\vskip-0.25in
\label{stats:audience_donors}
\end{table}

\para{Detailed targeting audiences} Although the ability to create Custom Audiences is only granted to advertisers with some history of running and paying for ads (the exact eligibility criteria are not publicly disclosed), all advertisers can specify their audience using detailed targeting (Section~\ref{subsec:targeting}).
We create a number of audiences this way, selecting a geographic region centered around a town and Facebook's inferred characterization such as ``Likely engagement with US political content (Conservative)'' and ``Likely engagement with US political content (Liberal)''. 
For some of the audiences we further narrowed the targeting by specifying additional required characteristics such as those who are, according to Facebook's characterization, ``interested in'' topics such as ``Donald Trump for President'', ``Make America Great Again'', ``Bernie Sanders'', or ``Elizabeth Warren''.
We aimed to approximately match the sizes of liberal and conservative audiences for each geographic region by adjusting the targeting radius around a chosen location until the Estimated Daily Reach matches.
The Appendix presents the details and size statistics for these audiences.

\subsection{Creating ad copies}\label{sec:ad-copies}
We ran three types of ads throughout our campaigns: (1) merchandise ads for candidates that link to the candidates' online campaign stores, (2) ``issues ads'' that have detailed content and that link to the candidates' websites, and (3) ``neutral'' political ads that simply encourage users to vote and link to generic voting information websites.\footnote{\url{https://www.usa.gov/election} and \url{https://www.usa.gov/register-to-vote}}
The majority of ads we ran were replicated from real ads run by official political campaigns obtained from the Facebook Ad Library~\cite{FacebookAdLibrary}. 
Ads for Bernie Sanders' merchandise store were the only exception, as---unlike the other campaigns in question---the Bernie Sanders campaign had not advertised merchandise on Facebook; we created the ad creative for this ad.
Whenever the replicated ad was written in the first person, we changed it to be a third person reference to the name of the candidate (as we were not running the ads {\em as} the campaign itself).
Examples of the ad copies of types (1) and (2) that we ran are presented in Figure~\ref{fig:ads}.

\subsection{Isolating role of content}\label{sec:servers}
Most of our ads link directly to either a candidate's official website or generic voting information websites.
In one of the experiments, however, we wanted to isolate the effect that the content of the advertised website has on the delivery skew, while keeping the users' reactions to the ad (such as possible Likes, comments, or reactions) constant.
We found that during ad creation, Facebook would automatically crawl the destination link as part of the ad review and classification process.
We develop a methodology that would use this feature to create ads that look like they have the same content to users, but different content to Facebook.

\begin{figure}[t!]
\centering
\includegraphics[width=.75\columnwidth]{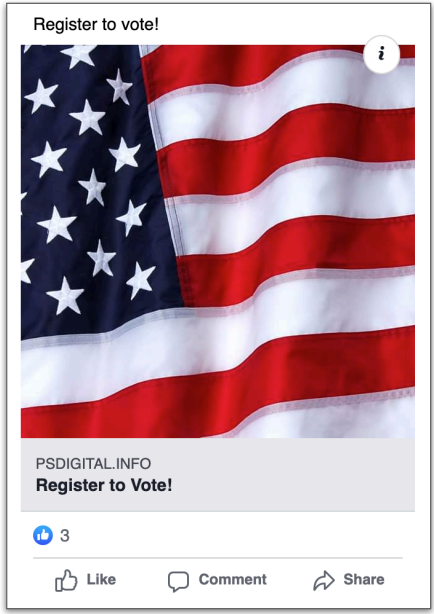}
\caption{Ads that have a destination link to our webserver (\texttt{psdigital.info}), which serves HTML from the candidate webpages to requests from Facebook's IP addresses, but redirects all other traffic to a generic voting information site.  The ads look identical to users, but different to Facebook.}
\label{fig:ads_self_hosted}
\vskip-0.15in
\end{figure}

To this end, we created a generic ad with a call to register to vote, a picture of the American flag, and a link to a nondescript domain: \texttt{psdigital.info} (see Figure~\ref{fig:ads_self_hosted}).
We created three copies of this ad, with each copy having a destination link to a {\em different page} under that domain.
We configured our web server to deliver a different response for requests for these pages based on the IP address of the requestor.
If the requestor was a Facebook-owned\footnote{We determined Facebook IP addresses by using the IP address blocks advertised by Autonomous Systems numbers owned by Facebook.} IP address, we served a copy of the HTML\footnote{Only the HTML code was served from our server; we modified the HTML so that  all images, JavaScript, and stylesheets would be downloaded from the corresponding official websites.} from the official Trump campaign website, the official Sanders campaign website, or a generic voting information website,\footnote{\url{https://www.usa.gov/register-to-vote}} depending on the particular page under our domain requested.
Otherwise, if the requestor was from any other IP address, the user would be immediately redirected to the generic voting information website.
In this way, all three ads would appear identical to users (and those users would all be brought to the same voting information site if they clicked on the ad), but Facebook's algorithm believed they linked to pages with different political content.

\subsection{Collecting performance statistics}
As mentioned in Section~\ref{subsec:reporting}, Facebook provides semi-live statistics on how the ad is delivering.
Once an ad starts running, we query Facebook every five minutes in order to get these statistics over the lifetime of the ad.
For ads where we use Custom Audiences with DMAs as a proxy for political leaning, we request these delivery statistics be broken down by DMA.

\subsection{Statistical analysis}
The core questions in this work revolve around comparisons of the fractions of Democrats (or Republicans) among the users exposed to two ads that differ in their content.
The comparison process consists of two steps and is based on previous work~\cite{ali-2019-discrimination}.

\textit{First}, we estimate the fraction of Democrats in each ad and the 99\% confidence interval around that estimate as shown in Equation~(\ref{eq:single_99}):
\begin{equation}
\begin{aligned}
L.L. &= \frac{\hat{p}+\frac{z^2_{\alpha/{2}}}{2n}-z_{\alpha/2}\sqrt{\frac{\hat{p}(1-\hat{p})}{n}+\frac{{z^2_{\alpha/2}}}{4n^2}}}{1+{z^2_{\alpha/2}}/n},\\
U.L. &= \frac{\hat{p}+\frac{z^2_{\alpha/{2}}}{2n}+z_{\alpha/2}\sqrt{\frac{\hat{p}(1-\hat{p})}{n}+\frac{{z^2_{\alpha/2}}}{4n^2}}}{1+{z^2_{\alpha/2}}/n},
\end{aligned}
\label{eq:single_99}
\end{equation}
where $L.L.$ is the lower confidence limit, $U.L.$ is the upper confidence limit, $\hat{p}$ is the observed fraction of Democrats in the audience, $n$ is the total size of the audience exposed to the ad. 
To obtain the 99\% interval we set $z_{\alpha/{2}}=2.576$.

\textit{Second}, we compare whether the fractions in two scenarios are statistically significantly different.
If their confidence intervals do not overlap (easily judged visually from the subsequent figures), the difference is statistically significant.
If the intervals do overlap, we need to perform a difference of proportion test as shown in Equation~(\ref{eq:zscore}):
\begin{equation}
Z = \frac{(\hat{p_1}-\hat{p_2})-0}{\sqrt{\hat{p}(1-\hat{p})(\frac{1}{n_1}+\frac{1}{n_2})}}
\label{eq:zscore}
\end{equation}
where $\hat{p_1}$ and $\hat{p_2}$ are the fractions of Democrats in the two audiences, $n_1$ and $n_2$ are the total sizes of these audiences, and $\hat{p}$ is the fraction of Democrats in the two audiences combined.
If the resulting $Z$-score is above $2.576$ (corresponding to 99\% confidence) the difference in proportion is statistically significant.

\section{Experiments}\label{sec:experiments}
We now present the detailed set-ups and results of our experiments.
Recall that our aim is to study both (a) whether the content of a political campaign's ad could lead to skew in delivery along political lines, and, if so, (b) whether a political campaign can successfully reach users who Facebook believes are not aligned with the campaign's views.
In the two subsections below, we address each of these questions in turn, before discussing the implications and limitations of our study.

\subsection{Ad content and skew}\label{subsec:contentskew}
We begin by examining whether the content of an ad can lead to skew in delivery along political lines.

\para{Voter records}
Similar to methodology of prior work~\cite{ali-2019-discrimination} for studying skews along race and gender, we use the Custom Audiences CA$_A$, CA$_B$, CA$_C$, and CA$_D$ described in Table~\ref{stats:audience_voters} that are based on voter records.
These audiences are designed so that asking Facebook to report delivery statistics by DMA serves as a proxy for obtaining delivery statistics by political affiliation.

We create three ad creatives: one taken from the official Donald Trump campaign, another from the Bernie Sanders campaign (both found in Facebook's Ad Library~\cite{FacebookAdLibrary}, shown in Figure~\ref{fig:ads} and linking to the respective campaign's web site), and a ``neutral'' political ad that simply encourages users to vote and links to a generic election website.\footnote{\url{https://www.usa.gov/election}}
We then run one copy of each ad targeting each of the four Custom Audiences, for a total of 12 individual ads.
Our ads are run with a daily budget of \$20 per ad set and use the objective ``Traffic" and optimization ``Link Clicks" (Section~\ref{subsec:objective}) as in prior work~\cite{ali-2019-discrimination}.

\begin{figure}[t!]
\centering
\includegraphics[width=1\linewidth]{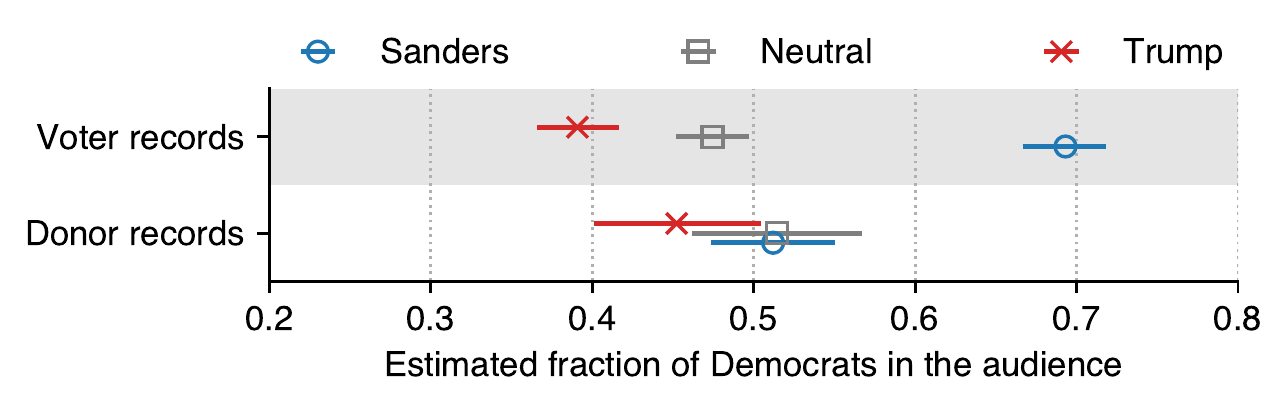}
\caption{The estimated fraction of Democrats who were shown our ads, targeting both registered voters in North Carolina and political donor records. In the case of voter records, the ad delivery to Democrats ranges from approximately 69\% for Sanders' ad to only 39\% for the Trump's ad.  In the case of donor records, we do not see statistically significant differences in ad delivery.}
\label{fig:nc_skew}
\end{figure}

Figure~\ref{fig:nc_skew} (top row) presents the overall delivery statistics for these three ads, with the delivery statistics of all four instances of each ad aggregated together. 
We can immediately observe significant differences in delivery:  The neutral ad delivers to 47\% Democrats, while the Trump ad delivers to less than 40\% Democrats. 
The Sanders ad, on the other hand, delivers to almost 70\% Democrats.
Note that this difference in delivery is despite the fact that all ads are run from the same ad account, at the same time, targeting the same audiences, and using the same goal, bidding strategy, and budget; {\em the only difference between them is the content and destination link of the ad}.

\para{Donor records}
Having observed that delivery skew along political lines can occur due to the content of the ad, we next turn to examine whether that skew is amplified if we choose users who recently engaged with politics.
In particular, we examine whether recent {\em donors} to political campaigns are estimated by Facebook to have greater relevance for our ads, when compared to users who are simply registered as Democratic or Republican voters.
Thus, we use our Custom Audiences of donor records (CA$_E$ and CA$_F$, described in Table~\ref{stats:audience_donors}); however, due to the limited size of the donor record databases, we are only able to run our experiment on two, and not four, audiences.
Thus, we run six ads in the same manner as the experiment we just described.

The results of this experiment are presented in Figure~\ref{fig:nc_skew} (bottom row).
Surprisingly, we do not find statistically significant differences in the ad delivery between the three ads targeting political donors. 
While we can only speculate as to why we observe a skew with voter records but not with donor records, the absence of a skew for the donor record audiences might suggest that Facebook does not have sufficient information about these users to do accurate relevance estimation for political ads. 
\begin{figure}[t!]
\centering
\includegraphics[width=1\linewidth]{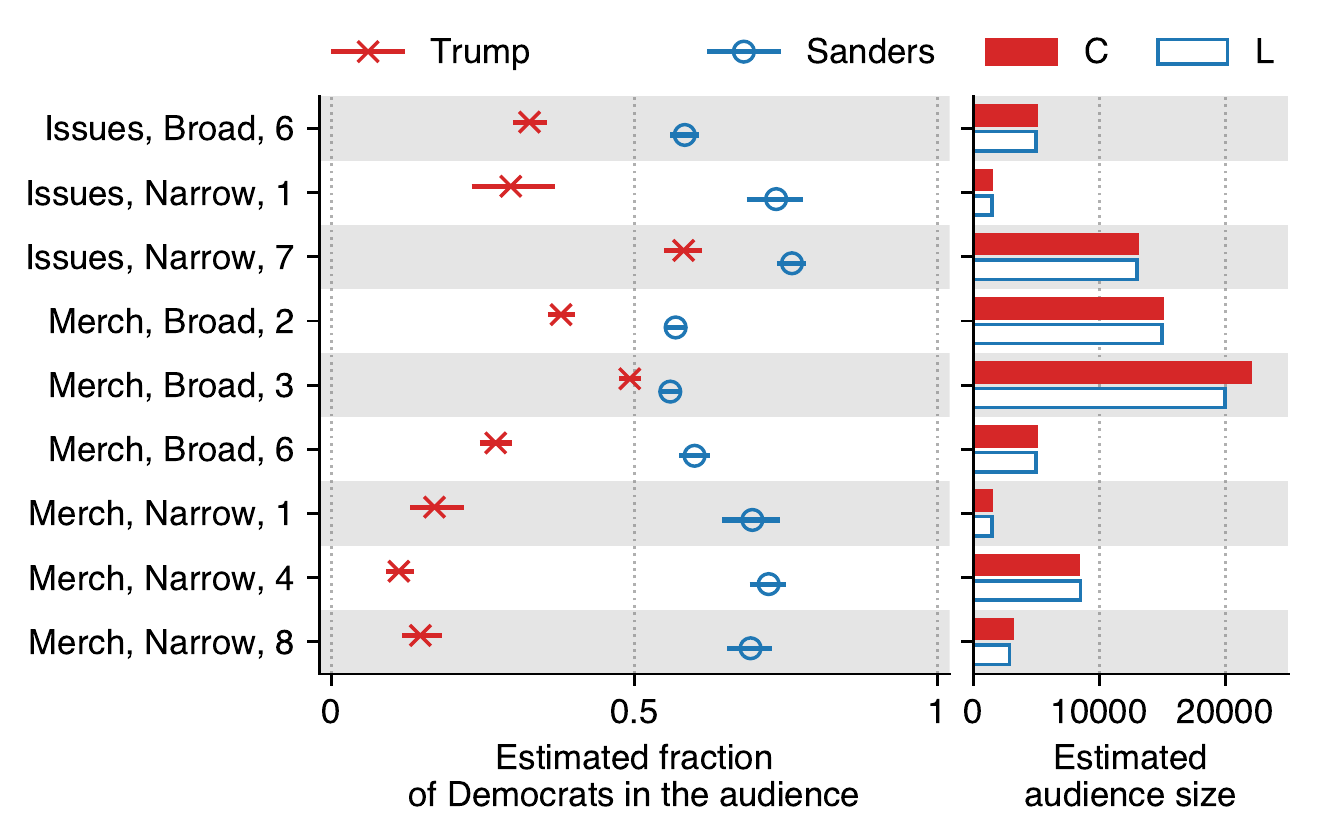}
\caption{We ran merchandise and issue ads with two levels of targeting specificity (Broad: users with ``Likely engagement with US political content (Conservative)'' or ``... (Liberal)"; Narrow: additional detailed targeting for inferred interest in Donald Trump or Bernie Sanders), and targeting different regions (1:\,Celina, OH; 2:\,Dutchess, NY; 3:\,Lorain, OH; 4:\,Macclenny, FL; 5:\,McCormick, SC; 6:\,Richlands, VA; 7:\,Saginaw, MI; 8:\,Slinger, WI). In all cases, Sanders' ads deliver to a larger fraction of Democrats than Trump ads even though they are targeting the same audiences at the same time using the same budgets. The effect is more pronounced for smaller audiences (compare, for example Merch, Broad, 3 and Merch, Broad, 6).}
\label{fig:all_reach}
\end{figure}

\para{Detailed targeting}
To further explore the role of Facebook's use of inferences about its users in delivery and its impact on political ad skew, we next use audiences where we {\em know} that Facebook has inferred the political affiliation of its users.
We do so using detailed targeting (Section~\ref{subsec:targeting}), selecting attributes ``Likely engagement with US political content (Conservative)'' for one audience and ``Likely engagement with US political content (Liberal)'' for another.
As discussed in Section~\ref{sec:creating-audiences}, we then geographically limit our targeting to regions where we can ensure an approximately equal number of users in each audience (as shown by Facebook's audience size estimates).
Then, over a course of six hours we concurrently ran two ad copies, each to two audiences (a total of four ads): one Sanders ad targeting liberal users and another targeting conservative users, and one Trump ad targeting liberal users and another targeting conservative users.
For this, and all further experiments, we optimize for ``Reach'', not ``Traffic''.
To calculate the delivery skew of a politician's ads, we sum reach across the two audiences, and calculate the fraction of deliveries to the liberal audience.

The result of our first experiment is shown in the top row of Figure~\ref{fig:all_reach}.
We can immediately observe similar skews in delivery to the ones observed for voter records, with the content of the ad causing delivery skew along political lines. 
This indirectly suggests that our hypothesis for the reasons behind differences for voter vs donor records could have some merit.

Next, we explore this finding in depth, varying three aspects of our experiment:
\begin{packed_enumerate}
\item{} The size of the audience, as reported by Facebook's Estimated Daily Reach,
\item{} The ``specificity'' of the audience (narrowing the detailed targeting further by attributes such as users' inferred interest in ``Donald Trump for President'' or ``Bernie Sanders'' according to Facebook), and
\item{} The specific topic of the ad (adding ads that advertise small campaign-branded merchandise that users can purchase, as shown in Figure~\ref{fig:ads}).
\end{packed_enumerate}
The results for these experiments are shown in Figure~\ref{fig:all_reach} (remaining rows), with each row representing a separate experiment.
Experiments described as ``issues'' are run with the first two ad creatives from Figure~\ref{fig:ads} and ``merch'' ads correspond to the third and fourth creative in Figure~\ref{fig:ads}.

We make a number of observations from this experiment.
{\em First}, we observe statistically significant skews in ad delivery along political lines for {\em all} of our ad configurations.
This suggests that such skew is a pervasive property of Facebook's ad delivery system.
{\em Second}, we observe that the skews tend to be less pronounced when the ads are targeting larger audiences (more than 10,000 daily active users).
While we do not know the underlying cause of this phenomenon, we hypothesize that the larger audiences provide the platform with a big enough pool of users to afford ``relevant'' users regardless of their inferred political leaning.
On the other hand, we suspect that when running our ads with smaller audiences, Facebook ``exhausts'' the (small) subset of users in the non-aligned audience (e.g., Sanders advertising to a conservative audience) for whom Facebook believes the ad is, in fact, relevant, and thus pauses or raises the price for delivery, but continues the delivery among the aligned audience.
We explore this hypothesis in more detail in the next experiment.

Overall, our findings strongly support our hypothesis that the content of an ad could lead to skew in its delivery along political lines whenever the platform has enough information (or thinks it has enough information) about the political leanings of the users being targeted, and that the skew is due to ad delivery optimization algorithms run by Facebook, rather than to other factors. 
As discussed in the introduction, this has profound implications for political advertisers, users, and society.

\begin{figure}[t!]
\centering
\includegraphics[width=1\linewidth]{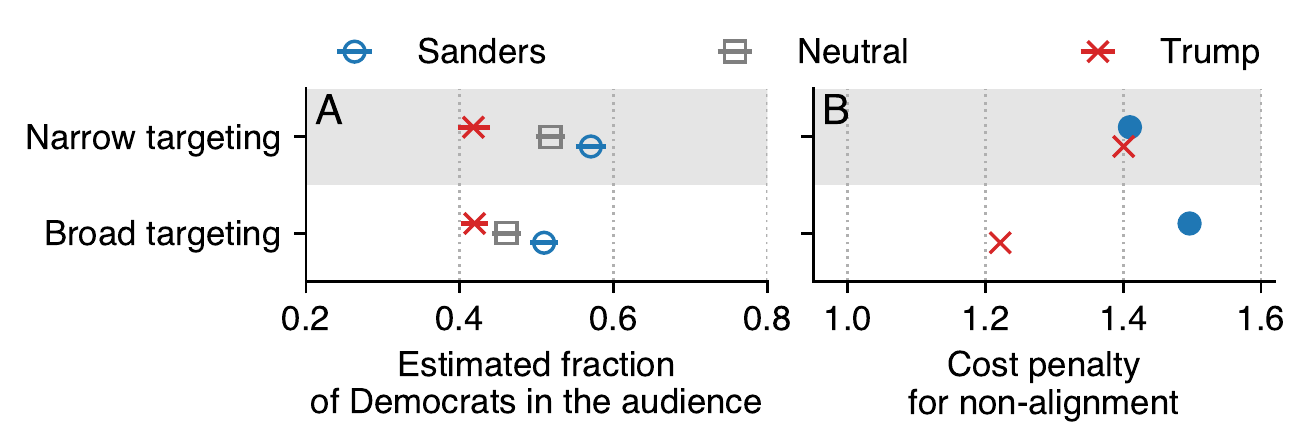}
\caption{Delivery statistics for ads that look identical to users, but appear partisan to the Facebook classification mechanism. (A) The skew in delivery is consistent with that observed in visibly distinct ads. (B) There is a financial penalty for trying to show an ad that Facebook deems non-aligned; reaching the same number of people in the same audience is up to 1.5 times more expensive.}
\label{fig:american_flag}
\end{figure}

\begin{figure*}[t!]
\centering
\includegraphics[width=1\linewidth]{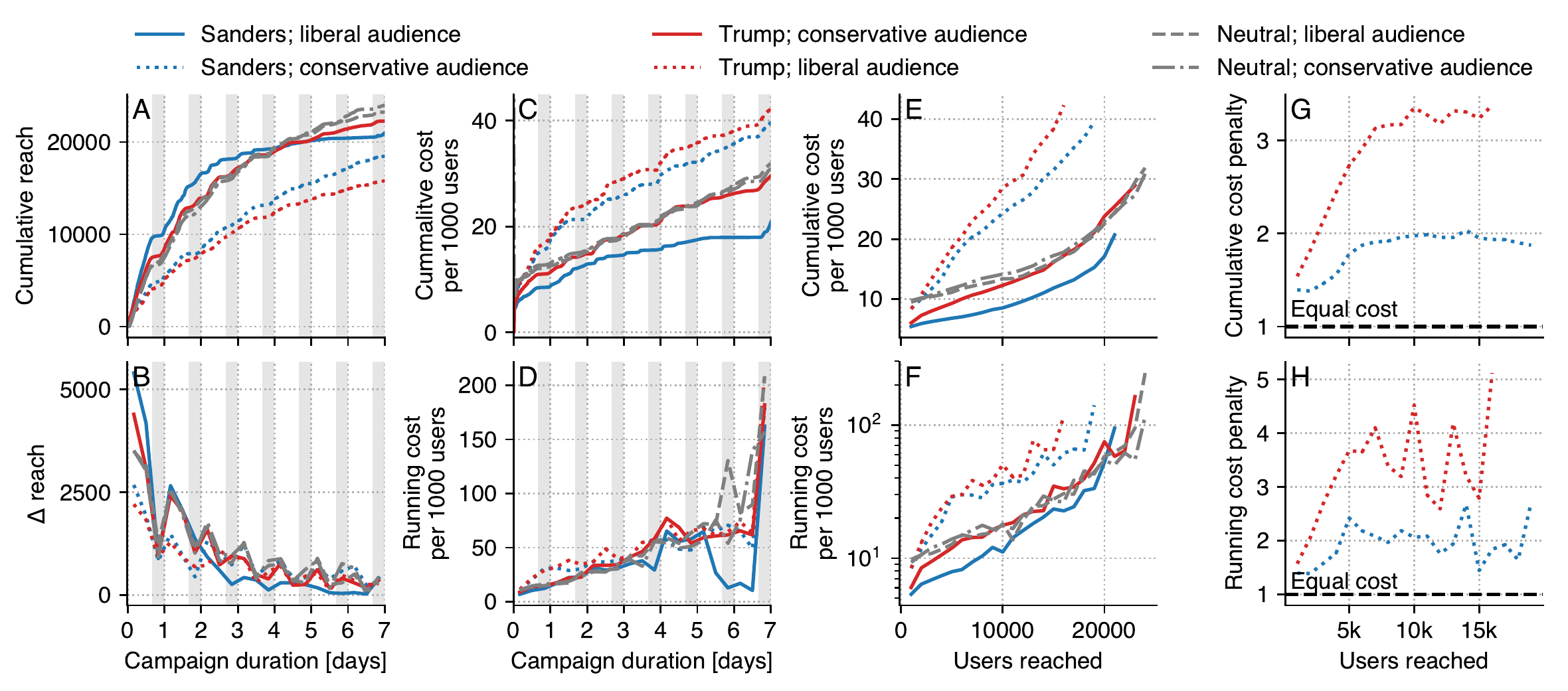}
\caption{Ads for a political campaign deliver to more users and for a lower cost if the targeted users have the same partisanship. A and B - the delivery rates are the highest in the beginning of the ad runtime and for aligned audiences. C and D - the cost of reaching non-aligned audiences is higher, especially in the beginning of the experiment. E and F - the more people have already seen the ad, the more expensive it becomes to show it to even more people; that growth is log-linear (see F). G and H show the ratio between the cost of a political campaign advertising to a non-aligned audience and their competitor advertising to the same audience.}
\label{fig:nc_long_campaigns}
\end{figure*}

\subsection{Longitudinal delivery}
We now explore what happens if a political campaign aims to reach users who Facebook believes are not aligned with the campaign's views.
Specifically, we ``force'' the Facebook ad platform to consider showing ads to all users in the political advertiser's targeting set, including users for whom Facebook may believe the ad is not relevant.
We do so in two steps: {\em First}, on a small audience we measure whether skew appears even if the ads look the same to users but differently to Facebook. 
{\em Second}, we run near-copies of real ads of political campaigns on a larger audience to measure the total effect. 
In both cases, we configure the campaigns using the objective ``Reach'' and optimization ``Reach'', which enables us to tell Facebook to only show the ad once to each user each week, thus forcing the delivery mechanism to `exhaust' the audience rather than showing the ad to the same subset of users.

\para{Generic ads} 
We begin by applying the method described in Section~\ref{sec:servers}, setting up ads that look identical to Facebook users (Figure \ref{fig:ads_self_hosted}) and, when clicked, redirect the user to a governmental website with instructions to register to vote.
However, when visited by Facebook's web crawler, each ad's landing website shows different HTML: one serves Trump's campaign HTML, another Sanders', and a third a generic voting information site.
Since all ads look identical to users, any skew can only be attributed to Facebook's optimization based on the content of the linked website.

Each ad copy is targeted at four audiences: two that are Broad and two that are Narrow.
The Broad audiences target ``Likely engagement with political content (Liberal)'' and ``...Conservative'', and the Narrow audiences additionally target users with interest in Bernie Sanders and Donald Trump, respectively.
Table~\ref{stats:audience_saved} entries for Oxford, NC and Scranton, PA provide detailed audience parameters and size statistics for the Broad and Narrow audiences respectively.
Since we are attempting to reach everyone in the targeting set here, we set a higher budget \$40 per day for each ad. 
These ads were run for two days and did not fully exhaust the audiences.
The results of this experiment are presented in Figure~\ref{fig:american_flag}A.
Even though the users see the same ad in all three cases, and therefore their explicit or implicit reactions to them are not more different than chance, the delivery is still skewed according to what Facebook's crawler sees. 
We also present the price differentiation in Figure~\ref{fig:american_flag}B.
For a given audience, we measure how much it cost for the aligned ad to reach the same number of users as the non-aligned ad did.
For example, it cost 1.5$\times$ more for the ad linking to Sanders' campaign page (as perceived by Facebook) to reach the same number of users in the Broad conservative audience than the ad linking to Trump's campaign page.
Conversely, it cost 1.2$\times$ more for the ad linking to Trump's campaign page to reach the same number of people in the Broad liberal audience than the ad linking to Sanders' campaign page.

These results show that the contents of the destination link---and not users' reaction or engagement with the ad---play a role in Facebook's decision for skewed delivery and differential pricing.
An implication of this finding is that two campaigns running an ad about the same issue to the same target audience might reach different fractions of that audience and at different prices, only because the destination links are different.
This differential delivery and pricing may be particularly damaging for local political campaigns, where candidates may agree on some issues but not others.

\para{Real ads}
We now turn to explore how this effect plays out for real-world ads that differ in content and destination link.
In this experiment, we run three ads (Trump, Sanders, and neutral issue ads as before), each to two Narrow audiences over a period of seven days and with a daily budget of \$100 for each ad and audience combination.\footnote{Our total spend over the week ended up being \$4,228.19 distributed roughly equally among Sanders, Trump, and neutral ads.}
The ad copies are the first two presented in Figure~\ref{fig:ads}, and details about the target audiences are provided in Table~\ref{stats:audience_saved} (the audiences from Michigan and Wisconsin).
The conservative and liberal audiences were selected such that they had approximately the same daily active reach, and such that we expected our ads to have reached {\em almost everybody} in the audience by the end of the seven days.

The results of this experiment are presented in Figure~\ref{fig:nc_long_campaigns}.
We first focus on panel A, which shows the cumulative number of users reached over seven days (along with its derivative in panel B).
We can observe that the delivery increases rapidly for all ads during the first day and then slows down quickly.
However, we can observe two notable outliers in panel A:  the Trump ad targeting the liberal audience and the Sanders ad targeting the conservative audience.
Both of these non-aligned ads end up delivering to over 25\% {\em fewer} users than their aligned counterparts.
In other words, when the Trump ad is advertised to the conservative audience, it delivers to a total of 21,792 users; when the Sanders ad is run at the same time and targeted to the same conservative audience, it delivers to only 17,964 users. 
Note that this difference cannot be attributed to some unknown underlying difference between Facebook use between the users categorized as liberal and conservative by Facebook because the neutral ad delivers equally to liberals and conservatives, reaching approximately 23,000 users.

We turn to panel C, which shows the cumulative cost per thousand unique users to help explain why this is occurring.
We can immediately notice an increasing cost trend for all ads:  as the ads run longer, their cost increases substantially.
Presumably, this is because Facebook first delivers the ad to the ``cheaper'' users in the target audience before deciding to spend our budget on the more ``expensive" users.
However, we can observe that the non-aligned ads are again outliers here: both show a substantially {\em higher} cost per thousand users, a difference noticeable from the first day of the experiment.
By the end of the experiments, when the liberal ad is shown to the liberal audience, it is charged \$21 per thousand users; when the conservative ad is delivered to the same audience, it is charged over \$40 per thousand users.

Because the delivery rates slow down after the first day plots C and D make the growth of cost per 1,000 also appear to slow down.
Therefore, we turn to plots E and F, which show this growth as a function of the size of reached audience, rather than time.
We observe that the growth is rapid, and the running cost per 1,000 is growing exponentially as a function of the number of users reached.
Finally, in plots G and H we show that the ratio between the cost a political campaign pays to show their ad to the non-aligned audience and the cost of their competitor showing to the same audience is relatively stable, between 2:1 and 4:1.

Overall, Figure~\ref{fig:nc_long_campaigns} emphasizes three findings:
\textit{First}, the penalty for reaching a non-aligned audience remains at a relatively stable ratio between 2:1 and 4:1 as a function of audience already reached (see Figure~\ref{fig:nc_long_campaigns}H). 
\textit{Second}, that while the cost per thousand viewers grows linearly with time (Figure~\ref{fig:nc_long_campaigns}D), it grows super-linearly with the number of users already reached (Figure~\ref{fig:nc_long_campaigns}F).
For example, panel F shows that the cost of showing the Sanders ad to the first 1,000 liberal users (solid blue line) is approximately \$5, and the cost of reaching the first 1,000 conservative users with this ad is approximately \$10.
However, once 10,000 users in each of these audiences are already reached, reaching another thousand of liberal users costs approximately \$15 (a three-fold increase) and reaching another thousand of conservative users costs approximately \$37 (nearly a four-fold increase compared to the first thousand conservatives).
\textit{Third}, at the end of the experiment, both neutral ads and the two aligned partisan ads reached over 20,000 users, while the non-aligned ads reached significantly fewer.
This demonstrates the core phenomenon: it is cheaper and more effective for a political campaign to reach audiences that are politically aligned (as inferred by Facebook) with their agenda, and as the campaign progresses it becomes more expensive to reach additional viewers.

\begin{figure}[t!]
\centering
\includegraphics[width=1\linewidth]{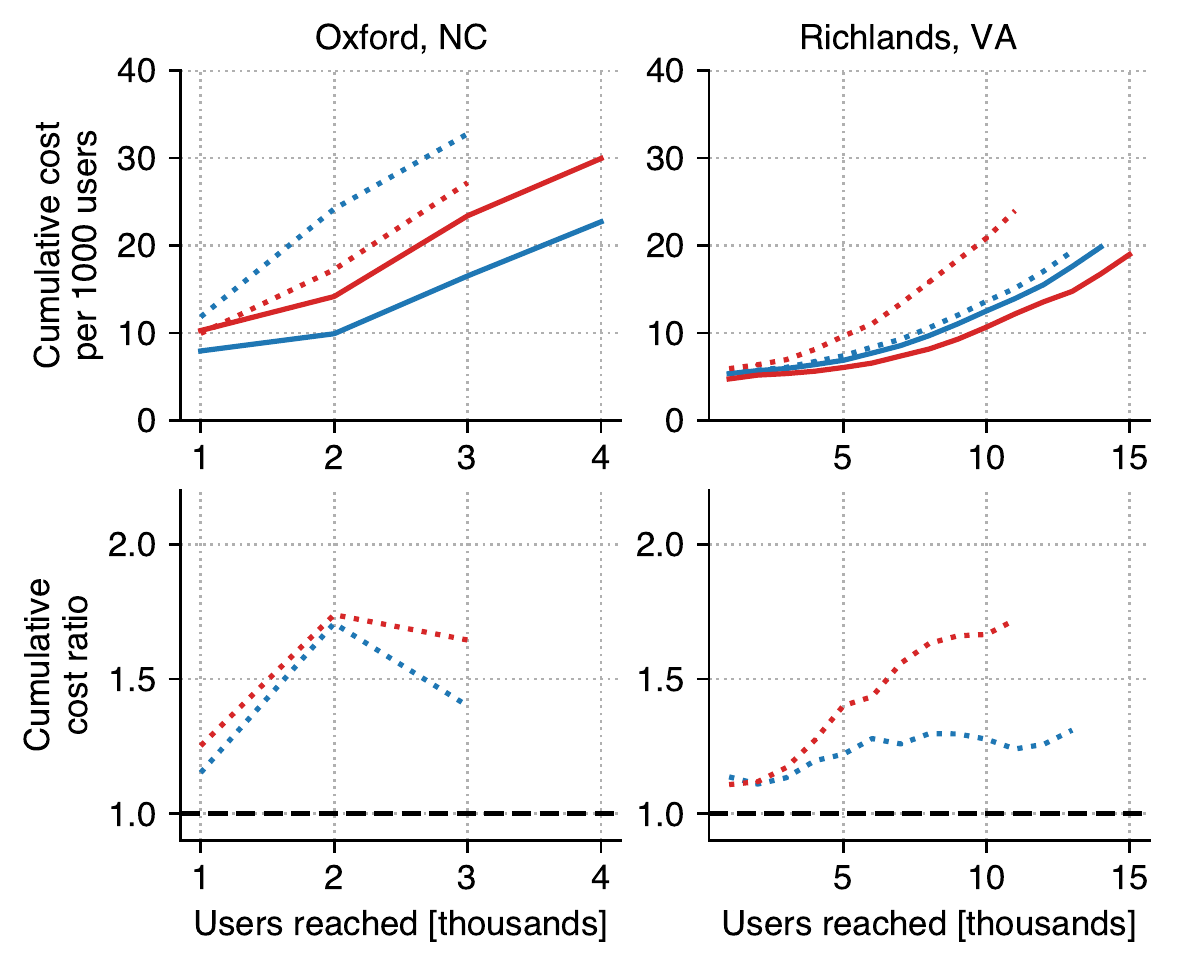}
\caption{The price penalty for showing to non-aligned audience is not just an effect of a particular location or targeting more ``extreme'' (Narrow) audiences. Findings from Figure~\ref{fig:nc_long_campaigns} hold also in other cities and with weaker targeting (here: only political alignment, without additional interests). We note that in some cities Trump's ads are cheaper for his aligned audience than Sanders' ads are for his.}
\label{fig:oxford-richlands}
\vskip-.15in
\end{figure}

Finally, we run a similar experiment targeting Broad audiences (\ie audiences without specific interest in candidates).
The results of this experiment are presented in Figure~\ref{fig:oxford-richlands}.
We find that the core phenomenon holds there as well, and note that in some cases, the conservative audience is cheaper for Trump than the liberal audience is for Sanders. 
\subsection{Limitations}\label{sec:limitations}
We now discuss limitations of our study and briefly mention the steps we took to mitigate them when possible. 
Controlling for all possible variables that may affect political ad delivery is beyond the scope and financial capabilities of our work, and is better suited to be performed by Facebook itself or by an independent third-party auditor that would be granted broader data and algorithms' access than what is available through the ad interface. 
Similarly, it is important to note that we can only report on delivery skew that we observed for our own ads; we {\em cannot} draw any conclusions about how political ads in general (or all ads run by a particular campaign) are delivered.
Nonetheless, the fact that we observe strong and statistically significant effects in our small set of ads suggests that the potential negative outcomes for individuals, political campaigns, and society in the context of ad delivery optimization of political advertising are not mere hypotheticals and warrant further scrutiny (Section~\ref{sec:discussion}). 

\para{Role of advertiser's identity} We have repeated a subset of our experiments using another advertising account registered as an advertiser in the area of ``Social Issues, Elections, and Politics'' and linked to a Facebook page unrelated to the first. Our results were quantitatively and qualitatively similar. This suggests that the effects we observed were not tied to our particular advertising account.  Nevertheless, we do not make any statements about the extent to which the observed effects hold when run by real political campaigns with a more established history than ours.

\para{Role of budget} We also re-ran a subset of our experiments with varying lifetime budgets ranging from \$10--\$100 per campaign, and with a generous bid cap of \$10 in each auction. Our ads ran on consecutive weekdays at similar times; we observed qualitatively similar skews regardless of the budget. Although \$100 per ad may seem small compared with the total political ad spending, such ads are representative of practice: recent work~\cite{NyuPoliticalAdPaper} that analyzes data from Facebook's political ad archive has found that 82\% of all political ads spend less than \$100.

\para{Role of competition}
We ran each pair of campaigns targeting a particular audience representing two different political campaigns at the same time and with the same budget. 
Such a set-up is designed to ensure that both campaigns have the same users available for delivery (i.e., if run at different times, the skews could be attributed to different Facebook use patterns by liberals or conservatives) and both are experiencing the same competition from other advertisers (i.e., that it would not be the case that one campaign is under-performing because it happened to run at the same time that another large and wealthy advertiser was targeting those users, whereas another campaign avoided such a collision). 
Thus, running campaigns simultaneously is an effective strategy to isolate the effects of delivery optimization from other extraneous factors. However, to verify that the skews are not merely the effect of our ads competing with each other, we also re-ran a subset of campaigns separately. The qualitative and quantitative skew effects for those campaigns we similar.

\para{Audience sizes} We aimed to match our constructed liberal and conservative audiences in size as closely as possible, but the matches are inevitably imprecise as Facebook only provides \textit{estimates} of daily reach\footnote{\url{https://www.facebook.com/business/help/1691983057707189?helpref=faq_content}} rather than audience sizes.

\para{User engagement with our ads}
There are a number of ways users can engage with the ads we present, each of which potentially influences future delivery and pricing:
reactions (`like', `love', `haha', `wow',  `sad', `angry'), commenting, and sharing.
Facebook advertising interface reports all such engagements.
Additionally, Facebook might be collecting and using {\em telemetric information}; for example, how long each user spent looking at the ad. 
This telemetric information is not available to the advertisers (and thus, neither to us), but might still play a role in ad delivery optimization algorithms.

Some of our ads received reactions, comments, and re-shares from the users they were delivered to. 
We note four important, related observations, that emphasize that our findings about skew in delivery and differential pricing are not merely a function of the ad delivery algorithm's use of user engagement.
{\em First}, we observe consistent skew and price differences in ads that look identical to users, yet trick Facebook into classifying them as partisan (Figure~\ref{fig:american_flag}).
Users do not react differently to ads that appear identical, and, therefore, the entire observed difference can be attributed to Facebook's pre-delivery classification (and some random effects).
{\em Second}, we observe consistent skew in delivery of ads that had virtually no engagement since they were run on small budgets and only for a few hours, as shown in Figure~\ref{fig:all_reach}.  
{\em Third}, longitudinal ads with neutral content proved less engaging than either aligned or not-aligned ads, yet they eventually reached larger audiences and at lower prices (Figure~\ref{fig:nc_long_campaigns}).
Specifically, non-aligned ads were shared at higher rates than neutral ads (0.34\% vs 0.03\% for conservative audience, 0.19\% vs 0.05\% for the liberal audience). 
This leads us to believe that the relatively lower costs of aligned ads compared to non-aligned ads do not stem from ``free'' exposures originating from re-shares.
{\em Finally}, we do find a negative correlation between the fraction of positive reactions (``like'' and ``love'') among all reactions and the price in the longitudinal ads with $\rho=-0.91$, $p_{val}=0.01$. 
Taken together, our work demonstrates although the skew in delivery as well as differential pricing can be further amplified during the course of delivery by users' reactions, the primary reason stems from Facebook's ad delivery optimization's use of classification of an ad and its landing page content.

We leave a more precise quantification of the influence of users' interactions with the ads on Facebook's ad delivery and pricing algorithms to future work.

\section{Discussion}\label{sec:discussion}
Our findings suggest that Facebook is wielding significant power over political discourse through its ad delivery algorithms without public accountability or scrutiny.

\para{Implications}
\textit{First,} Facebook limits political advertisers' ability to reach audiences that do not share those advertisers' political views in ways that are significantly different from traditional broadcast media. 
The existence and extent of this skew may not be apparent to advertisers and varies based on their ad's message and the destination link used by the campaign. 
For example, a campaign targeting a certain geographic region might reasonably expect to reach an audience whose political views are representative of users in the region. 
To discover otherwise would require careful research, as we have demonstrated in this study. 
Furthermore, the strength of delivery skews vary for campaigns of different political leanings and targeting different populations, making digital advertising inequitable for political campaigns with identical budgets.

\textit{Second}, recent moves to restrict political advertisers' targeting options~\cite{TwitterPoliticalAdsBan,GooglePoliticalAdsPolicyUpdate,FacebookPoliticalAdsPolicyChange}, although valuable from a user privacy perspective~\cite{korolova-2011-ads, faizullabhoy-2018-fbattacks, speicher-2018-targeted}, might be undermined by the operation of ad delivery algorithms, and even give companies like Facebook {\em more} control over selecting which users see which political messages. 
This selection occurs without the users' or political advertisers' knowledge or control. 
Moreover, these selection choices are likely to be aligned with Facebook's business interests, but not necessarily with important societal goals.

\textit{Third}, today, researchers, regulators, and campaigns lack access to algorithms and data required for a more thorough study of ad delivery skews and their likely impacts. 
In particular, although much has already been said about the inadequacy of current ad transparency tools provided by ad platforms~\cite{MozillaInadequateAdArchive,KnightPublicInterest,UpturnFacebook}, our work draws attention to the need to expand these efforts to account for ad delivery algorithms as well.

\para{Policy analysis}
Today, U.S. law cannot do much, if anything, to {\em directly} change how ad platforms deliver political ads. For the foreseeable future, it is likely that the primary regulator of digital political advertising will not be the government, but rather ad platforms themselves.

The U.S. Congress has addressed conceptually similar "ad delivery issues" in the past, albeit in a different domain. 
For example, the Federal Communications Commission (FCC) enforces the so-called Equal-Time Rule~\cite{EqualTimeRule}, which originated in 1927 in response to worries that broadcast licensees could unduly influence the outcome of elections. 
The rule requires that licensees make air time available to all candidates for the same office on equivalent terms.  
However, the rule only applies to broadcast licensees, and has only narrowly survived constitutional scrutiny in part because it implicates government interests in managing limited broadcast spectrum~\cite{CbsRevoking}.

Prevailing interpretations of the First Amendment are likely to block efforts to extend the logic of the Equal-Time Rule to digital advertising platforms, which are not regulated like broadcast licensees. 
As an initial matter, the First Amendment strongly protects political speech, and generally tolerates only narrowly-tailored government regulations~\cite{wood-2017-fool}.
This protection is so strong that legal scholars cannot even be confident that lighter-touch kinds of regulations---for example, a requirement that social media users be entitled to opt-in to micro-targeted political advertising---would survive constitutional scrutiny. 
Moreover, the Supreme Court recently declared that ``the creation and dissemination of information" constitutes speech under the First Amendment~\cite{SorellIms}.
This reasoning, which might expand the ``commercial free speech'' rights of companies, creates some uncertainty about the government's ability to restrict corporations' use of data in the context of digital advertising. 

Looking ahead, it is clear that government regulation of digital political advertising is on firmest legal footing when it requires disclosure about who is speaking to whom, when, and about what~\cite{wood-2017-fool}. 
Accordingly, Congress and the FEC can consider transparency requirements that will enable detailed auditing and research about ad targeting and the delivery of political ads.

\para{Mitigations}
As an initial data, the public and the campaign managers need more information about the operation of ad delivery algorithms and their real-world effects. 
Ad platforms could increase transparency around political ads (including key metrics such as targeting criteria, detailed ad metadata, ad budgets, and campaign objectives) to enable further study of the effects of ad targeting and delivery. 
And they could provide access to and insight into the ad delivery algorithms themselves (including those involved in running the auction, relevance measurement and estimation, and bid and budget allocation on advertisers' behalf), allowing third parties greater ability to study and audit their performance and effect on political discourse. 
Without these and similar steps, policymakers and the public will be unable to formulate appropriate responses.

Ad platforms could also disable delivery optimization for political content, or a least allow advertisers to do so. 
They could also introduce more nuanced user-facing controls for political content delivery and expand public ad archives to make them more accessible and usable by everyone. 

Finally, we call on ad platforms to acknowledge the central role they play in the delivery of political ads, and to collaborate with other key stakeholders---including researchers, political campaigns, journalists, law, policy and political philosophy scholars---to address that role when it is not aligned with public interests.

\section*{Acknowledgements}
We would like to thank Dean Eckles and David Lazer for their invaluable insights.
We are also extremely grateful to the participants and organizers of the REAL ML workshop for their encouragement and constructive feedback.
This work was done, in part, while Aleksandra Korolova was visiting the Simons Institute for the Theory of Computing, where she benefited from supportive feedback from participants of the Privacy and Fairness programs, and particularly, from the suggestions of Amos Beimel and Kobbi Nissim.
This work was funded in part by a grant from the Data Transparency Lab, NSF grants CNS-1616234, CNS-1916020, and CNS-1916153, and Mozilla Research Grant 2019H1.

\section*{Errata}
\textbf{v3}: Clarified the discussion on efficacy of personality-based targeting.

\noindent\textbf{v2}: We clarified the optimization goal used in the experiments: {\em Traffic} in experiments with donors and registered voters and {\em Reach} in all other experiments. 

\balance
\newcommand{\etalchar}[1]{$^{#1}$}

\newpage
\appendix
\section*{Appendix}
\label{sec:appendix}

\para{Audiences from Facebook's targeting attributes.} Table~\ref{stats:audience_saved} shows the geographical location, targeting parameters, and estimated potential reach given by Facebook. In each case, the geographical diameter and targeting specificity is tweaked to have roughly equal sized liberal and conservative audiences.

\begin{table}[ht!]
\centering{}
\scriptsize
\resizebox{\columnwidth}{!}{
\begin{tabular}{p{2cm}p{4cm}cr}
\toprule
\ra{2}
{\bf Location} & {\bf Targeting} & \phantom{a} & {\bf Size} \\
\midrule
\makecell*[{{p{2cm}}}]{Celina, OH \\ (+25 mi)} & 
\makecell*[{{p{4cm}}}]{\textbf{Engagement}: liberal\\
\textbf{Interests}: Bernie Sanders}
&& 1,500 \\
\makecell*[{{p{2cm}}}]{Celina, OH \\ (+21 mi)} & 
\makecell*[{{p{4cm}}}]{\textbf{Engagement}: conservative\\
\textbf{Interests}: Donald Trump for President}
&& 1,400 \\
\makecell*[{{p{2cm}}}]{Dutchess\\ County, NY} & 
\makecell*[{{p{4cm}}}]{\textbf{Engagement}: liberal\\}
&& 15,000 \\
\makecell*[{{p{2cm}}}]{Dutchess\\ County, NY} & 
\makecell*[{{p{4cm}}}]{\textbf{Engagement}: conservative\\}
&& 15,000 \\
\makecell*[{{p{2cm}}}]{Loraine, OH\\ (+13 mi)} & 
\makecell*[{{p{4cm}}}]{\textbf{Engagement}: liberal}
&& 20,000 \\
\makecell*[{{p{2cm}}}]{Loraine, OH\\ (+10 mi)} & 
\textbf{Engagement}: conservative
&& 22,000 \\
\makecell*[{{p{2cm}}}]{Macclenny, FL\\ (+24 mi)} & 
\makecell*[{{p{4cm}}}]{\textbf{Engagement}: liberal\\
\textbf{Interests}: Bernie Sanders}
&& 8,500 \\
\makecell*[{{p{2cm}}}]{Macclenny, FL\\ (+30 mi)} & 
\makecell*[{{p{4cm}}}]{\textbf{Engagement}: conservative\\
\textbf{Interests}: Donald Trump for President}
&& 8,300 \\
\makecell*[{{p{2cm}}}]{McCormick, SC\\ (+20 mi)} & 
\makecell*[{{p{4cm}}}]{\textbf{Engagement}: liberal}
&& 3,000 \\
\makecell*[{{p{2cm}}}]{McCormick, SC\\ (+17 mi)} & 
\makecell*[{{p{4cm}}}]{\textbf{Engagement}: conservative}
&& 3,400 \\
\makecell*[{{p{2cm}}}]{Richlands, VA\\ (+34 mi), VA} & 
\makecell*[{{p{4cm}}}]{\textbf{Engagement}: liberal}
&& 5,000 \\
\makecell*[{{p{2cm}}}]{Richlands, VA\\ (+10 mi), VA} & 
\makecell*[{{p{4cm}}}]{\textbf{Engagement}: conservative}
&& 5,000 \\
\makecell*[{{p{2cm}}}]{Saginaw, MI\\ (+10 mi)} & 
\makecell*[{{p{4cm}}}]{\textbf{Engagement}: liberal}
&& 13,000 \\
\makecell*[{{p{2cm}}}]{Saginaw, MI\\ (+10 mi)} & 
\makecell*[{{p{4cm}}}]{\textbf{Engagement}: conservative}
&& 13,000 \\
\makecell*[{{p{2cm}}}]{Slinger, WI\\ (+21 mi)} & 
\makecell*[{{p{4cm}}}]{\textbf{Engagement}: liberal\\
\textbf{Interests}: Bernie Sanders, U.S. Senator Bernie Sanders}
&& 2,900 \\
\makecell*[{{p{2cm}}}]{Slinger, WI\\ (+24 mi)} & 
\makecell*[{{p{4cm}}}]{\textbf{Engagement}: conservative\\
\textbf{Interests}: Donald Trump for President}
&& 3,100 \\
Michigan & 
\makecell*[{{p{4cm}}}]{\textbf{Engagement}: liberal\\
\textbf{Interests}: Democratic Party (United States), Bernie Sanders, U.S. Senator Bernie Sanders, Joe Biden, Barack Obama}
&& 34,000 \\
\makecell*[{{p{2cm}}}]{Wisconsin\\ and Michigan} & 
\makecell*[{{p{4cm}}}]{\textbf{Engagement}: conservative\\
\textbf{Interests}: Donald Trump for President, Republican Party (United States), Make America Great Again or Mike Pence}
&&  38,000\\
\makecell*[{{p{2cm}}}]{Oxford, NC\\ (+12 mi)} & 
\makecell*[{{p{4cm}}}]{\textbf{Engagement}: liberal}
&&  3,000\\
\makecell*[{{p{2cm}}}]{Oxford, NC\\ (+10 mi)} & 
\makecell*[{{p{4cm}}}]{\textbf{Engagement}: conservative}
&&  3,600\\
\makecell*[{{p{2cm}}}]{Scranton, PA\\ (+45 mi)} & 
\makecell*[{{p{4cm}}}]{\textbf{Engagement}: liberal\\
\textbf{Interests}: Bernie Sanders, Democratic Party (United States), U.S. Senator Bernie Sanders, Barack Obama, Joe Biden}
&&  3,000\\
\makecell*[{{p{2cm}}}]{Scranton, PA\\ (+50 mi)} & 
\makecell*[{{p{4cm}}}]{\textbf{Engagement}: conservative\\
\textbf{Interests}: Donald Trump for President, Make America Great Again}
&&  3,200\\
\bottomrule
\end{tabular}
} 
 \caption{Overview of the audiences created using Facebook's inferred interests. The targeting parameters are chosen so that the number of conservative and liberal users in audiences corresponding to a particular location are roughly equal. Interests are combined together to narrow audiences down (i.e. with a logical AND), except for the Wisconsin and Michigan audience, where one OR clause is used.}
 \label{stats:audience_saved}
 \end{table}

\end{document}